%% file: 00-main.tex
\documentclass[sigconf,preprint,review,screen,authordraft,nonacm,review=false,timestamp=false]{acmart}

\usepackage[frozencache,cachedir=.]{minted}
\usepackage{enumitem}
\usepackage{caption}
\usepackage{subcaption}

\usepackage{mdframed}

\usepackage{color}
\usepackage{soul}
\usepackage{cleveref}
\usepackage{nth}

\usepackage{pifont}
\newcommand{\cmark}{\ding{51}}%

\crefformat{section}{\S#2#1#3} 
\crefformat{subsection}{\S#2#1#3}
\crefformat{subsubsection}{\S#2#1#3}

\usepackage[colorinlistoftodos]{todonotes}

\newcommand{\contribution}{distributed execution indexing}
\newcommand{\Contribution}{Distributed Execution Indexing}
\newcommand{\ContributionAbbr}{DEI}
\newcommand{\ThreeMB}{\textsc{3MileBeach}}
\newcommand{\Filibuster}{\textsc{Filibuster}}
\newcommand{\SFIT}{Service-level Fault Injection Testing}

\newcommand{\SFITAbbr}{SFIT}

\begin{document}

\author{Christopher S. Meiklejohn}
\affiliation{%
  \institution{Carnegie Mellon University}
  \city{Pittsburgh, PA}
  \country{USA}
}
\email{cmeiklej@cs.cmu.edu}

\author{Rohan Padhye}
\affiliation{%
  \institution{Carnegie Mellon University}
  \city{Pittsburgh, PA}
  \country{USA}
}
\email{rohanpadhye@cmu.edu}

\author{Heather Miller}
\affiliation{%
  \institution{Carnegie Mellon University}
  \city{Pittsburgh, PA}
  \country{USA}
}
\email{heather.miller@cs.cmu.edu}

\renewcommand{\shortauthors}{Meiklejohn \textit{et al.}}




\title{Distributed Execution Indexing}
\subtitle{Work-in-progress Report}

\begin{abstract}
This work-in-progress report presents both the design and partial evaluation of distributed execution indexing, a technique for microservice applications that precisely identifies dynamic instances of inter-service remote procedure calls (RPCs).
Such an indexing scheme is critical for request-level fault injection techniques, which aim to automatically find failure-handling bugs in microservice applications.
Distributed execution indexes enable granular specification of request-level faults, while also establishing a correspondence between inter-service RPCs across multiple executions, as is required to perform a systematic search of the fault space.
In this paper, we formally define the general concept of a distributed execution index, which can be parameterized on different ways of identifying an RPC in a single service. 
We identify an instantiation that maintains precision in the presence of a variety of program structure complexities such as loops, function indirection, and concurrency with scheduling nondeterminism. 
We demonstrate that this particular instantiation addresses gaps in the state-of-the-art in request-level fault injection and show that they are all special cases of distributed execution indexing.	
We discuss the implementation challenges and provide an implementation of distributed execution indexing as an extension of \Filibuster{}, a resilience testing tool for microservice applications for the Java programming language, which supports fault injection for gRPC and HTTP.
\end{abstract}

\maketitle

\input{01-introduction}
\input{02-fault-injection}
\input{03-dei}
\input{04-implementation}
\input{05-evaluation}
\input{06-conclusion}

\bibliographystyle{ACM-Reference-Format}
\bibliography{main}

\end{document}

%% file: 01-introduction.tex
\section{Introduction}
\label{sec:introduction}

Resilience engineering is a process performed by many software and site reliability engineers today to ensure that their application is resilient to faults.  
One specific class of faults that are a concern for the developers of microservice applications are \textit{partial failures}: where the unavailability of one or more dependent services can render an entire application unusable. 
One way that engineers anticipate this inevitable partial failure is by implementing \textit{fallbacks}: where, in the event of a failure of one service, a different service can compensate for that failure.
For example, by replacing content that is unavailable due to failure with different content on a user's homepage, as done by Netflix~\cite{10.1145/3472883.3487005}.
Therefore, an important component of resilience engineering is ensuring that this fallback behavior works correctly.

There are several different approaches for ensuring fallback behavior works correctly.
For example, chaos engineering~\cite{chaos-monkey, 10.1109/ICSE-SEIP.2019.00012, 10.1145/2987550.2987555} identifies these issues using stochastic, coarse-grained fault injection performed in the live, production environment that affects actual end-user traffic.
Complementary to chaos engineering, request-level fault injection (RLFI)~\cite{10.1145/3472883.3486986} enables resilience engineering by introducing failures at the level of an individual remote procedure call (RPC). 
RLFI enables \SFIT{} (\SFITAbbr{})~\cite{10.1145/3472883.3487005} a systematic search technique for microservice applications that exhaustively covers the space of inter-service RPC failures.
SFIT starts with an existing functional test that exercises application behavior when no faults are present.  
SFIT then re-executes these tests repeatedly, each time choosing one or more RPCs to fail until all faults and combinations of RPCs have been tested.

RLFI-based techniques require the ability to uniquely identify a dynamic instance of an RPC in order to target it for fault injection.
\SFIT{} requires that these identifiers are also deterministic across multiple test executions to support systematic search.
Existing RLFI implementations~\cite{10.1145/3472883.3486986, 10.1145/3472883.3487005} use a combination of RPC signatures and invocation counts per signature to do this identification.
However, they fail to account for common programming patterns found in microservice applications: for example, loops containing RPCs, branching control flow statements that contain RPCs, and the use of concurrency primitives for invoking RPCs, where scheduling nondeterminism is possible.  
These limitations may result in an \textit{unsound} analysis, as faults may be injected on an incorrect inter-service RPC.
In the specific case of exhaustive search, these limitations may result in an \textit{incomplete} analysis, where valid faults are not explored.

In this paper, we present \textit{\contribution{}} (\ContributionAbbr{}), a technique for precisely identifying an inter-service RPC uniquely for RLFI, and in the case of \SFIT{}, deterministically across multiple test executions.

To see why such an indexing scheme is important and challenging to devise, consider an example of a booking service for cinemas.  As part of the process of retrieving a users's reservations, the \emph{users} service contacts the \emph{bookings} service and the \emph{movies} service to retrieve information on both the users's bookings and detailed information on each booked movie.
A trivial way to identify RPC invocations would be to use the source and destination service name, as a pair of strings, allowing the system to distinguish between the two different RPCs: (\emph{users}, \emph{bookings}) vs. (\emph{users}, \emph{movies}). Of course, this will not be sufficient if there are more than one RPCs between the same pair of services in a given end-to-end execution; e.g. the \emph{users} service may invoke more than one method of the \emph{bookings} service, and perhaps repeat this for multiple users.

Contemporary RLFI systems uses RPC signatures to identify a particular RPC instance to inject a failure on. Signatures include the destination service, the invoked method name, and it's parameters. 
\ThreeMB{}~\cite{10.1145/3472883.3486986} additionally accumulates the entire causal history of RPCs that are invoked during an execution, and uses invocation counts for each identifier can be then used to distinguish between multiple RPC invocations with the same signature.
Further extending our cinema example, this would alter our identifiers to also contain the list of all previously invoked RPCs: if the \emph{movies} service was invoked after \emph{bookings}, we can consider the identifier of the RPC to \emph{movies} to contain the identifier of the preceding RPC to \emph{bookings} as well. As such, the identifiers are path-sensitive.
While this approach accounts for multiple invocations with the same signature, it fails to account for branching conditional statements where an RPC with the same signature is invoked on multiple branches of the conditional. \Filibuster{}~\cite{10.1145/3472883.3487005} also uses RPC signatures for identification, but keeps track of the call stack at the time of invocation, as well as an invocation counter associated with each stack state.
This serves to distinguish RPCs in loops and branching conditional statements. However, \Filibuster{} is designed specifically for single-threaded Python microservices that communicate using HTTP for RPC; its identification scheme fails to account for concurrency and scheduling nondeterminism, where multiple RPCs may be invoked between the same pair of services with the same call stack state at the same time, and their ordering cannot be controlled.

The problem of scheduling nondeterminism manifests itself in RLFI through the permutation of RPCs: concurrent RPCs with the same signature can be executed in different orders across multiple executions, thereby permuting the identifiers associated with each. One solution for deterministic identifier assignment is to explicitly control thread scheduling. Unfortunately, this is not a feasible approach for large, microservice applications because of two core issues. First, these applications are typically implemented in microservice frameworks, and rely on RPC frameworks, where threads are reused for performance. Second, these applications are typically implemented in multiple languages across many different services, where using a centralized thread scheduler is not practical.

\begin{mdframed}[skipabove=0.1cm, skipbelow=0.1cm, nobreak]
\noindent \textbf{Key observation and insight.}
Our solution leverages the following \textit{key observation}: while microservice applications may issue concurrent RPCs with the same signature, these concurrent RPCs will rarely contain the same payload: the precise arguments provided to that RPC.
Therefore, our \textit{key insight} is that the inclusion of the RPC payload in each RPC's identifier enables the deterministic assignment of identifiers to each RPC without requiring control of thread creation or thread scheduler.
We name our formulation \emph{distributed execution indexing}, as it extends the concept of {execution indexing}~\cite{10.1145/1379022.1375611} to distributed systems.
\end{mdframed}

Empirically validating our key observation about concurrent RPCs is not a straightforward task, due to the lack of research corpora that contain microservice applications.  
In fact, recent academic work focused on the construction of a microservice application corpus~\cite{10.1145/3472883.3487005} that contains resilience bugs \textit{do not contain a single example} where an individual service issues concurrent RPCs.
To address this limitation, we recently established a working relationship with an industrial partner that runs a large microservice application with over 500 services in order to validate our technique at scale.
We plan to use this partnership to perform the empirical validation needed to justify our key observation and key insight.

Finally, we present a partial synthetic evaluation of \ContributionAbbr{} that supports function indirection, looping, branching control flow and concurrency, performed using the \Filibuster{} application corpus.  
In the process of this evaluation, we extend the \Filibuster{} corpus with several new valuable examples demonstrating problematic patterns that will be valuable for future researchers in microservice resilience testing.  
Along with this contribution to the \Filibuster{} corpus, we provide an open-source implementation of \ContributionAbbr{} and an extension of \Filibuster{} that supports both gRPC \& HTTP, in Java. 

The contributions of this paper are the following:

\begin{itemize}
    \item \textbf{a formal description of \ContributionAbbr{}.} (\cref{sec:dei}) \\
    We present a formal definition of \ContributionAbbr{}, a technique for identifying inter-service RPCs in microservice applications where these identifiers establish a correspondence across multiple executions. 
    We demonstrate that existing RLFI techniques use special cases of \ContributionAbbr{}.
    
    \item \textbf{an implementation of \ContributionAbbr{}}. (\cref{sec:implementation}) \\
    We provide an open-source implementation of \ContributionAbbr{} implemented in Java for the JVM, integrated directly into a new \Filibuster{} client library for Java that supports both gRPC and HTTP (using Armeria.)  
    We discuss challenges in both the implementation of \ContributionAbbr{} and its integration into \Filibuster{}.
    
    \item \textbf{an evaluation of \ContributionAbbr{}}. (\cref{sec:evaluation}) \\
    We present a preliminary, synthetic evaluation of \ContributionAbbr{} using the \Filibuster{} open-source microservice application corpus.  
    We also contribute two new examples, that demonstrate problematic programming patterns for existing RLFI techniques, to this open-source microservice application corpus.
\end{itemize}

%% file: 02-fault-injection.tex
\section{Microservices and Fault Injection}
\label{sec:fi}
\begin{figure}
  \includegraphics[width=\linewidth]{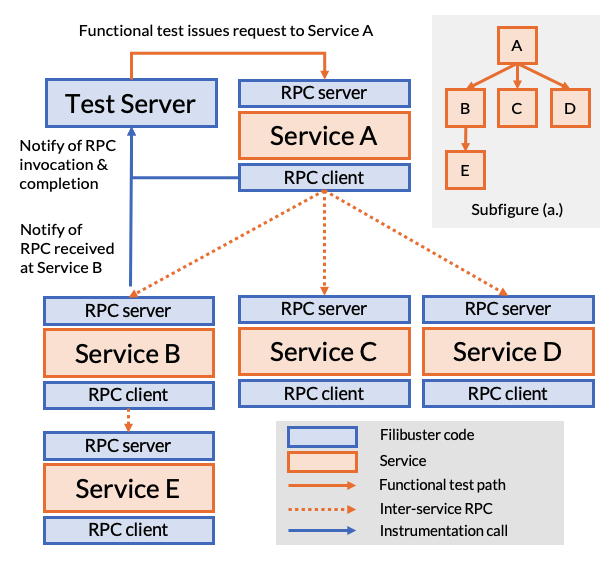}
  \caption{Example microservice application with instrumentation that enables \SFITAbbr{}.}
  \label{fig:fi:example-application}
\end{figure}

In order to demonstrate the complexity of microservice applications and the fault-injection methodology that is required to guarantee resilience, we consider the microservice application presented in Figure~\ref{fig:fi:example-application}a.
In Figure~\ref{fig:fi:example-application}, there are five services.  
Service A receives requests from the end user and issues RPCs to B, C, and D before returning a response to the end user; B issues an RPC to E before returning a response to A. 
As the developers of these applications assume that any one of A's direct dependencies may be unavailable at any point, they design each service in a manner where default responses for each RPC are used in the event that any one of the RPCs should fail; they do the same for Service B's direct dependency as well.
The question that remains is: \textit{does this fallback behavior functions as expected?}

One approach for verifying that this fallback behavior works as expected is through the use of test mocks.
This approach requires that developers, for each point in the fault space, write a functional test to exercise the application under fault along with the mock necessary to simulate that fault.  
The fault space is large: at a minimum, developers must first consider each individual RPC and all of the exceptions that can be thrown by that RPC; then, they must explore all combinations.

More often than not, these tests are not written.
We believe this to be a result of the effort required and the complexity involved in running an entire microservice application locally while mocking individual RPCs for failure.
Somewhat ironically, it is these very tests that are most important to system reliability: recent research investigating~\cite{186171} critical errors in open source distributed data systems has identified that the error handling code responsible for preventing these errors was either missing or never tested.

While distributed data systems are not microservice applications, in some ways they are simpler and easier to test: they contain fewer distinct services, as these services are typically deployed in replica sets, they are homogeneous in their implementation language, and usually have well-defined behavior under failure, as they typically implement a distributed protocol.  
 In contrast, microservice applications typically have hundreds of different services, all with different behavior and developed by independent teams, where a single service can rely on multiple, different distributed data stores for storage of application state.  
 Further complicating matters, behavior under failure may not be well-defined, as developers working on individual services may not realize a certain failure is possible when it is only triggered by a certain combination of simultaneous faults across different services.
 We believe that if developers are not writing these tests for distributed data systems, they most likely are not writing them for microservice applications either. 
 
\subsection{Service-level Fault Injection Testing}
\SFIT{} (\SFITAbbr{}) enables a systematic search for microservice applications using RLFI that exhaustively covers the space of inter-service RPC failures.

\paragraph{Static Analysis.} 
\SFITAbbr{} starts by identifying the faults that each service in an application should be tested for.
This constructs the fault space that will be exhaustively searched.
To do this, \SFITAbbr{} starts by statically analyzing each framework that is used to issue RPCs in order to determine the throwable exceptions at each call site. 
For example, both HTTP and gRPC frameworks throw exceptions when the remote host is unreachable.
Then, the implementation of each service is also analyzed to identify any response types, specific to that service, that are used to indicate failure.  
For example, an HTTP service can indicate failure through either 400 or 500-series error codes; for gRPC, specific error codes can indicate different failures such as \texttt{FAILED\_PRECONDITION}.

\paragraph{Instrumentation.} 
\SFITAbbr{} relies on instrumented versions of each RPC framework.
We depict this in Figure~\ref{fig:fi:example-application}.
This enables \SFITAbbr{} to identify RPC invocations, identify where those RPCs are received, propagate required metadata information between inter-service RPCs, and perform fault injection.
This is orchestrated by a centralized test server.

\paragraph{Dynamic Analysis.}
For a test oracle, \SFITAbbr{} uses an existing functional test suite for the microservice application.
For each functional test, \SFITAbbr{} will first run an initial execution where no faults will be injected.
It will then repeatedly re-execute the test, injecting different combinations of faults, until the fault space has been exhausted.
Using Figure~\ref{fig:fi:example-application} to demonstrate, the RPC between B and E would first be tested for all possible faults identified through our static analysis. 
Then, the RPCs between A and its direct dependencies B, C, D would be tested for those faults as well, including all possible combinations thereof.

\begin{mdframed}[skipabove=0.1cm, skipbelow=0.1cm, nobreak]
\cmark~The exhaustive search performed by this dynamic analysis requires that dynamic invocations of RPCs are identified both uniquely and deterministically across all test executions in order to identify when the exhaustive search is complete.
\end{mdframed}

\paragraph{Test Adaptation.}
As faults are injected, assertions in the test oracle will fail.  
This is \textit{expected}, as the test only encodes application behavior when no faults are present.
Therefore, developers will be prompted to use conditional assertions to encode the desired behavior under failure; effectively encoding derivations from the behavior under fault depending on where the fault as injected.

\paragraph{Dynamic Reduction.}
Finally, to keep the technique scalable and avoid executing redundant tests \SFITAbbr{} leverages a property of microservice applications called \textit{service encapsulation}.  
Service encapsulation states simply that if there are two services, Services X and Y, where X invokes RPCs on Service Y, any failure of Y is only visible to the callers of X as a failure or success of X itself: in this case, A is said to \textit{encapsulate} B.  
This property holds true as long as the two services (A) do not share state in the same database, (B) are deterministic in their responses for a given test, and (C) do not contain data dependencies on previous failures.
These properties are both derived from, and in line with, proper microservice design and testing tenets.
 
Using Figure~\ref{fig:fi:example-application} to demonstrate, the test execution where faults are injected in Service D and Service E simultaneously is considered redundant with this optimization.  
The reason for this is because Service B encapsulates Service E.  
The only way that Service A can encode conditional logic on the failure of Service D and Service E simultaneously failing is by using one of the responses from E to determine that.  
However, given that \SFITAbbr{} has already tested the execution where B and D have simultaneously failed, it has already observed this outcome.

\begin{mdframed}[skipabove=0.1cm, skipbelow=0.1cm, nobreak]
\cmark~Similar to the requirements of dynamic analysis, this optimization of the search also requires unique and deterministic identification of dynamic invocations of RPCs.
\end{mdframed}

%% file: 03-dei.tex
\section{\Contribution}
\label{sec:dei}
At the root of the \SFITAbbr{} analysis, is the need to uniquely (and deterministically) identify each RPC.  
For example, in Figure~\ref{fig:fi:example-application}, specific identification of the RPC issued between B and E. 
In order for \SFITAbbr{} to know when the systematic search is complete, this specific RPC --- between B and E --- must be identified the same across all test executions.  
While this may seem rather trivial, as it only requires the identification of a single edge in a microservice graph, it becomes more complex when multiple RPCs between the same pairs of services may exist in the same test execution.
The programming patterns that cause this behavior are rather commonplace: loops, branching, function indirection, and concurrency.

We need to define what it means for the \SFITAbbr{} analysis to be correct. 
In terms of correctness, we will concern ourselves with only the dynamic component of \SFITAbbr{}, as it is responsible for the assignment of unique (and deterministic) identifiers to each RPC.
As with all analyses, they must ideally be both sound and complete.
\textbf{Soundness} is violated when either a fault is injected on an RPC invocation where not intended or by the failure to inject a fault where intended. 
\textbf{Completeness} is violated when a required fault injection, for exhaustive search, is missed.

\subsection{Signatures Are Too Coarse-Grained}
Consider one simple way of identifying RPCs, as discussed in the introduction, the RPC's signature.
We formally define an RPC signature as follows:

\begin{definition}
\label{def:signature}
A \textit{signature} is a triple $(m, f, a)$ where

\begin{itemize}[]
    \item $m$ is the module or class name of the RPC stub;
    \item $f$ is the method or function name; and
    \item $a$ is the parameter names and types.
\end{itemize}

With gRPC, the class name and method map directly; parameters are the parameter types and names for the gRPC endpoint.
With HTTP, the URI and HTTP method can be combined to form the signature as it contains the target service, method name, and parameter names and types, which are assumed to be \texttt{String}.
\end{definition}

Let us see how the RPC signature is too coarse-grained to uniquely (and deterministically) identify an RPC and may result in an unsound or incomplete analysis.

\begin{figure}
\begin{minted}[obeytabs=true,tabsize=4,linenos,numbersep=-10pt,fontsize=\footnotesize]{python}
    @service_a.method("helloworld")
    def service_a_helloworld():
        hello = echo("Hello")
        world = echo("World")
        s = hello + " " + world
        return s
    
    def echo(s : String):
        try:
            res = rpc(service_b, "echo", s)
            log_success(res)
            return res
        except Exception as e:
            log_error(e)
            return s
        
    @service_b.method("echo")
    def service_b_echo(s : String):
        return s
\end{minted}
\caption{RPC signature alone cannot distinguish between the RPCs issued on lines 3 and 4; call stack or invocation count must be combined with signature.}
\label{fig:helloworld:fallbacks}
\end{figure}

Consider the example in Figure~\ref{fig:helloworld:fallbacks}.
In this example, we present a microservice application composed of 2 services written in pseudocode.
Service A exposes a single RPC endpoint, \texttt{helloworld}, which issues two RPCs to B's RPC endpoint, \texttt{echo}, before combining the responses and returning a response.
In the event that Service B is down, a default response is returned by the function wrapping the RPC, \texttt{echo}, on line 8.

In the case of the RPC invocation at line 10, the signature, would be composed of the target service name B, the method \texttt{echo}, and the parameter \texttt{(s,String)}.
In this application, the signature for both of the RPCs invoked by Service A, on lines 3 and 4, would be identical: $(\mathrm{\texttt{B}}, \mathrm{\texttt{echo}}, \mathrm{\texttt{(s,String)}})$. SFIT would not be able to distinguish between the first and second RPCs for systematic fault injection; that is, the RPC signature alone is too \textit{coarse-grained} for identifying a particular RPC. 

\subsection{Increasing Granularity: \\ Invocation Count or Call Stack}
One solution for resolving the issue where identical identifiers are assigned to different RPCs is to increase the granularity of the identifiers that we assign.
We examine two different ways that this could be accomplished and demonstrate that they must be used together. In the following discussion, since we're going beyond just signature-based identifiers, we assume (for ease of presentation and without loss of generality) that a service (say \texttt{A})) makes RPC invocations to only one other service (say \texttt{B}) and only a single RPC endpoint (\textit{e.g.} \texttt{echo}) per service. Thus, we use only the \textit{invoking service name} (\textit{e.g.,} \texttt{A}) as a shorthand for an outgoing RPC from \texttt{A} that stands in for the full signature which would contain the target service, method name, and parameters. 

\begin{enumerate}[]
    \item \textit{Invocation count.}  
    \ThreeMB{}~\cite{10.1145/3472883.3486986} and \Filibuster{}~\cite{10.1145/3472883.3487005} both keep track of the number of invocations for each RPC call site in order to distinguish multiple calls to the same call site.  
    In Figure~\ref{fig:helloworld:fallbacks}, the same RPC is invoked twice. We use the ``$|$'' symbol to indicate the invocation count of an RPC signature. For example, the identifiers $A|_1$ and $A |_2$ distinguish the \nth{1} and \nth{2} RPC invocations made from service \texttt{A} at lines 10.  
    
    \item \textit{Call stack.}
    Another approach is to increase the granularity of the identifier with some representation of the call stack.
    In Figure~\ref{fig:helloworld:fallbacks}, the RPC is invoked twice at line 10, however, with different calling contexts for the \texttt{echo} function (lines 3 and 4).  We use a \textit{superscript} to indicate the line number(s) corresponding to call stack at the time of invocation. For example, the two RPC invocations in Figure~\ref{fig:helloworld:fallbacks} can be distinguished by identifiers $A^{3,10}$ and 
    $A^{4,10}$.
\end{enumerate}

For the example in Figure~\ref{fig:helloworld:fallbacks}, either invocation count or call-stack based identification works to disambiguate the two RPCs. However, neither approach is sufficient on its own in general. A better approach is to use a \emph{combination} of invocation count \emph{and} calling contexts for identifying RPCs, e.g. $A^{3,10}|_1$, denoting the first invocation of RPC from A with the calling context $(3, 10)$.
To demonstrate the need for both these terms, we refer the reader to Figure~\ref{fig:helloworld:loops}.
In Figure~\ref{fig:helloworld:loops}, we present a different implementation for A; we assume that the implementation of B from Figure~\ref{fig:helloworld:fallbacks} is the same.  
In this example, A's RPC endpoint \texttt{helloworld} takes, as parameters, a list of \texttt{String}.
For each \texttt{String} that is provided, a RPC is invoked to B's \texttt{echo} endpoint.  
In the event that the RPC to B throws an exception, the remainder of the list traversal is aborted and a final RPC is made to B using a default value and that value returned by A.
When no exceptions are thrown, the aggregated results are joined and returned by A.

\begin{figure}[t]
\begin{minted}[obeytabs=true,tabsize=4,linenos,numbersep=-10pt,fontsize=\footnotesize]{python}
    @service_a.method("helloworld")
    def service_a_helloworld(ss : List[String]):
        rs = []
        failure = False
        
        for s in ss:
            try:
                r = rpc(service_b, "echo", s)
                rs.append(r)
            except Exception as e:
                failure = True
                break
                
        if failure:
            s = "Hello World"
            r = rpc(service_b, "echo", s)
            return r
        else:
            return rs.join(" ")
\end{minted}
\caption{Signature combined with invocation count insufficient in distinguishing \nth{2} iteration of loop from \nth{1} invocation of failure handler; signature combined with call stack insufficient in distinguishing loop iterations.}
\label{fig:helloworld:loops}
\end{figure}

Consider a functional test that invokes \texttt{helloworld} with a list containing two \texttt{String}s.
For simplicity, we assume that each RPC can only throw a single runtime exception.
Therefore, we must execute 5 different executions of the test to fully exhaust the fault space.
First, consider the execution where both loop iterations execute and all RPCs are successful, which we denote as a sequence of RPC invocations: $e_1: (A^8|_1, A^8|_2)$.
Next, we consider the executions where the RPC throws an exception, using the $\neg$ symbol to denote a failed RPC invocation. When a fault is injected in the \nth{2} iteration of the loop, there are two cases when the fallback RPC either completes successfully or fails: $e_2: (A^8|_1, \neg{A^8|_2}, A^{16}|_1) ,~e_3: (A^8|_1, \neg{A^8|_2}, \neg{A^{16}|_1})$.
Finally, we consider the executions where the RPC throws in the \nth{1} iteration and the fallback RPC either completes successfully or fails: $e_4: (\neg{A^8|_1}, A^{16}|_1),~e_5: ( \neg{A^8|_1}, \neg{A^{16}|_1})$.

Using this example and these test executions, we now examine why invocation count and call stack are, by themselves and in combination with the signature, insufficient for ensuring correctness based on our criteria.  
Therefore, they must be combined.

\begin{itemize}[]
\item \textbf{Invocation Count Alone is Insufficient.} 
Consider executions $e_1$ and $e_4$.   
Using this technique, $e_1: (A|_1, A|_2)$ and $e_4: (\neg{A|_1}, A|_2)$.
However, $A|_2$ in $e_1$ refers to the invocation at line 8 and $A|_2$ in $e_4$ refers to the invocation at line 16.
Therefore, to properly assign identifiers to these RPCs, we must increase the granularity to include the call stack that resulted in the RPC invocation.
\item \textbf{Call Stack Alone is Insufficient.}  
In $e_1$, both requests would be assigned the same identifier: $e_1: (A^8, A^8)$.
Therefore, to properly assign identifiers to these RPCs, we must increase the granularity to include the number of times each RPC invocation statement is reached.
\end{itemize}

\subsection{Increasing Granularity: Payload}
\label{sec:dei:payload}
While the addition of the call stack and invocation count to the RPCs signature are sufficient for distinguishing RPC invocations in the presence of loops and function indirection, they are not sufficient in the presence of concurrency and scheduling nondeterminism.

For example, consider Figure~\ref{fig:helloworld:async}, a modified version of Figure~\ref{fig:helloworld:loops}, where line 7 invokes an RPC using the \texttt{async} primitive and the results are \texttt{await}ed on line 11. 
In this example, the invoked RPCs execute concurrently and both their execution order is susceptible to \textit{scheduling nondeterminism}.   
This scheduling nondeterminism poses problems for an analysis like \SFITAbbr{}, where it may result in both an unsound or incomplete analysis.

Similar to before, we assume a functional test that invokes the \texttt{helloworld} RPC endpoint with two \texttt{String}s.  
For example, the first test execution that we run should read as follows: $e_1: (A^7|_1, A^7|_2)$: $A^7|_1$ is the RPC invoked in the \nth{1} iteration of the loop, where $A^7|_2$ is the RPC invoked in the \nth{2} iteration of the loop.
However, on repeated execution of this test through deterministic replay, or when performing exhaustive search, scheduling nondeterminism may result in the \nth{2} iteration of the loop being assigned $A^7|_1$, if the \nth{2} block happens to execute first.

\begin{figure}
\begin{minted}[escapeinside=||,obeytabs=true,tabsize=4,linenos,numbersep=-10pt,fontsize=\footnotesize]{python}
    @service_a.method("helloworld")
    def service_a_helloworld(ss : List[String]):
        rs = []

        for s in ss:
            r = |\colorbox{yellow}{async}| { 
                return rpc(service_b, "echo", s) 
            }
            rs.append(r)

        |\colorbox{yellow}{awaitAll}| rs
        return rs.join(" ")
\end{minted}
\caption{Scheduling nondeterminism can permute assignment of identifiers. In this case, $A^7|_1$, can refer to the RPC invocation from either the \nth{1} or \nth{2} loop iteration.}
\label{fig:helloworld:async}
\end{figure}

Model checkers for distributed systems~\cite{267763, 10.5555/2685048.2685080, 10.1145/3302424.3303986} also face the problem of scheduling nondeterminism.  
However, these model checkers were originally designed for identifying concurrency bugs before later being extended for failure testing (\textit{e.g.,} message omission) and therefore rely on control of the thread scheduler.
Even previous work on using execution indexes in multithreaded programs to detect deadlocks (\textit{e.g.,} \textsc{DeadlockFuzzer}~\cite{10.1145/1543135.1542489}) relies on specialized compilation when testing, for scheduler control.

Controlling the scheduler is an unrealistic for large, microservice applications where \textit{a.)} they may not be able to run all services on a single machine during testing, and where \textit{b.)} services are implemented in a number of different languages.
Therefore, we set out to identify a solution that did not require control of the thread scheduler.
We examine three different ways that this could be accomplished and demonstrate that none are sufficient.

\begin{enumerate}[]
    \item \textit{Cloning per block.}
    One approach is to \textit{clone} the state that is used to generate identifiers for each asynchronous block.
    This would ensure that each block would count invocations for each RPC signature, and associated call stack, independently.
    However, this approach does not work.
    In Figure~\ref{fig:helloworld:async}, this technique would result in identical identifiers for each of the RPCs executed during the loop: $(A^7|_1, A^7|_1)$.
    \item \textit{Encode thread creation.} 
    \textsc{DeadlockFuzzer}~\cite{10.1145/1543135.1542489}, a system for detecting deadlocks in concurrent programs using execution indexes, proposed an approach where thread creation is included in the identifier.  
    This approach does not work in the case of asynchronous blocks, as they may execute on an existing thread pool provided by the system or framework where the threads have already been created.
    \item \textit{Cloning per thread.}
    If we were to follow this line of thinking, we could also \textit{clone} the state that is used to generate the identifiers for each thread.
    This does not work either.
    In Figure~\ref{fig:helloworld:async}, scheduling nondeterminism may cause two of the RPCs to execute on a single thread in one execution $(A^7|_1, A^7|_2)$ and on two different threads in a subsequent execution: $(A^7|_1, A^7|_1)$.
\end{enumerate}

The approach that we arrived at as most practical stems from our \textit{key observation} about microservice applications: while these applications may issue concurrent RPCs with the same signature, these concurrent RPCs will rarely contain the same payload: the precise argument values supplied at invocation time.

Therefore, our \textit{key insight} is that, through the inclusion of the payload in each RPC's identifier, identifiers will be assigned deterministically without requiring control of thread creation or the thread scheduler.  
To achieve this, we \textit{share} the state used to derive identifiers across all threads that are used to execute concurrent code by reference.  
We refer to this as the \emph{invocation payload}.

\begin{definition}
\label{def:invocation-payload}
The \textit{invocation payload} $p$ for an RPC with $n$ parameters
is a sequence
$(k_1, v_1)(k_2, v_2)...(k_n, v_n)$
such that for each $i$ in $[1, n]$, the term $k_i$ is the $i$-th argument's name and $v_i$ is the $i$-th argument's value.

For gRPC, these are the precise argument values at invocation time.
For HTTP, these are the combination of query-string arguments and request body.
\end{definition}

In Figure~\ref{fig:helloworld:async}, and assuming the concrete argument provided to the function is the list \texttt{["Hello", "World"]}, we represent the execution: $e_1: (A(\mathrm{\texttt{(s,Hello)}})^7|_1, ~A(\mathrm{\texttt{(s,World)}})^7|_1)$.  
It is important to note that the invocation count in both of these identifiers is $1$, as it considers both the call stack and payload together.  
This ensures deterministic assignment regardless of scheduling nondeterminism.

We can use the combination of RPC signature, the calling context, and the invocation payload to create a dynamic \emph{invocation signature}, defined as follows:

\begin{definition}
\label{def:invocation-signature}
The \textit{invocation signature} for an RPC invocation is a triple $(s, p, t)$, usually denoted as $s(p)^t$, where:
\begin{itemize}[]
    \item $s$ is the signature of the RPC;
    \item $p$ is the invocation payload of the RPC; and 
    \item $t$ is a representation of the call stack of the RPC.
\end{itemize}

Thus, the notation $s(p)^t|_k$ refers to the $k$-th invocation of an RPC with invocation signature $s(p)^t$.
\end{definition}

An important point to note is that while RPC signatures (Definition~\ref{def:signature}) can be statically determined, the invocation signatures (Definition~\ref{def:invocation-signature}) are determined only based on observed executions.

While we have framed our presentation in this section using \texttt{async}/\texttt{await}, many other concurrency primitives (\textit{e.g.,} futures, coroutines) exist that have the same challenges.   
We believe our technique extends to all of them.

\subsection{Increasing Granularity: \\ Path to Currently Invoked RPC}
In Figure~\ref{fig:helloworld:path-encoding}, we present another variation on our \texttt{helloworld} microservice application.
Similar to Figure~\ref{fig:helloworld:loops}, Service A receives a list of \texttt{String}s, invokes an RPC on Service B for each member in the list, and accumulates the result.  
In the event of an exception, a placeholder value is accumulated and the failure is recorded.  
The recorded failures are then iterated in a retry loop and, if successful, the value replaces the placeholder.
Different from Figure~\ref{fig:helloworld:loops}, Service B invokes an RPC on a third service, Service C, and decorates the response somehow before returning a response to Service A.

\begin{figure}[t]
\begin{minted}[obeytabs=true,tabsize=4,linenos,numbersep=-10pt,fontsize=\footnotesize]{python}
    @service_a.method("helloworld")
    def service_a_helloworld(ss : List[String]):
        rs = []
        failure = False
        failures = []
        
        for s in ss:
            try:
                r = rpc(service_b, "echo", s)
                rs.append(r)
            except Exception as e:
                failure = True
                failures.append(len(rs) - 1, s)
                rs.append("")
                
        if failure:
            for (i, s) in failures:
                try:
                    r = rpc(service_b, "echo", s)
                    rs[i] = r
                except Exception as e:
                    pass
            
        return rs.join(" ")
            
    @service_b.method("echo")
    def service_b_decorate_echo(s : String):
        try:
            r = rpc(service_c, "echo", s)
            return r
        except Exception as e:
            return s
        
    @service_c.method("echo")
    def service_c_echo(s : String):
        return s
\end{minted}
\caption{RPC signature, when extended with invocation count and call stack, is insufficient when RPC invocation is triggered by different incoming RPC requests.}
\label{fig:helloworld:path-encoding}
\end{figure}

We assume a functional test that issues an RPC to A containing a list of two \texttt{String}s: \texttt{Hello} and \texttt{World}.
We abbreviate these to $H$ and $W$ and omit the parameter name \texttt{s} in their invocation signatures.
Using the technique from the previous section, the execution where the list iteration completes and no faults are injected is: 
$e_1: (A(H)^9|_1, B(H)^{29}|_1, A(W)^9|_1, B(W)^{29}|_1)$.
For each iteration, A issues an RPC from line 9 to B; when B receives the RPC from A, it issues an RPC to C from line 29. 

Now, let us consider the test execution where a fault is injected on the RPC in the \nth{2} iteration of the loop represented as follows: 
$e_2: (A(H)^9|_1, B(H)^{29}|_1, \neg{A(W)^9|_1}, A(W)^{19}|_1, B(W)^{29}|_1)$.
As before, during the \nth{1} iteration of the loop, Service A issues an RPC to Service B at line 9; Service B then issues an RPC to Service C at line 29. 
When we reach the \nth{2} iteration of the loop, a fault is injected for the RPC from Service A to Service B.  
Then, the failure condition is met and a subsequent RPC is issued from Service A to Service B on line 19; Service B then issues an RPC to Service C on line 29 before returning a response.

The issue we experience in this example is that the RPC identified by $B(W)^{29}|_1$ in test execution $e_1$ is \textit{not} the same as the RPC identified by $B(W)^{29}|_1$ in test execution $e_2$.
In execution $e_1$, the RPC from Service B to Service C at line 29 is caused by the RPC issued by Service A on line 9.
In execution $e_2$, the RPC from Service B to Service C at line 29 is caused by the RPC issued by Service A on line 19.  
These are not the same, even though they issue the same RPC with the same arguments and payload.
They represent distinct call sites in different parts of the code: one is part of the normal operation of the RPC endpoint where no failure occurs, and one represents error handling code that needs to be tested to ensure correct operation of the application under failure.
Therefore, associating the same identifier to these RPCs results in both unsound and incomplete behavior: either, the injection of faults on the incorrect RPC or the failure to explore the fault space during exhaustive search.  

To resolve this issue, we need to include the path of RPC invocations that resulted in the current RPC, as this information is not captured by the call stack.  
To achieve this, we accumulate a list of identifiers as we invoke RPC messages from service to service as part of handling a received RPC invocation: for example, $[A(W)^9|_1 :: B(W)^{29}|_1])$ to indicate that the \nth{1} invocation of invocation signature $B(W)^{29}$ occurred as a result of the \nth{1} invocation of invocation signature $A(W)^9$.
We can reformulate test executions $e_1$ and $e_2$ as follows:

\begin{itemize}[]
    \item $e_1: ([A(H)^9|_1], [A(H)^9|_1 :: B(H)^{29}|_1], $ \\
          $~~~~~~~~~[A(W)^9|_1], [A(W)^9|_1 :: B(W)^{29}|_1])$ \\
          The RPC invocations from A to B on line 9 are denoted with the prefixes $A(H)^9|_1$ and $A(W)^9|_1$ to include the enclosing RPC from A.
  \item $e_2: ([A(H)^9|_1], [A(H)^9|_1 :: B(H)^{29}|_1], [\neg{A(W)^9|_1}],$ \\
        $~~~~~~~~~[A(W)^{19}|_1], [A(W)^{19}|_1 :: B(W)^{29}|_1])$ \\
        The RPC invocation from B to C on line 29 is prefixed by $A(W)^{19}|_1$ which distinguishes it from the \nth{2} RPC in execution $e_1$ from A to B on line 9 that triggered the RPC from B to C on line 29.
\end{itemize}

\begin{definition}
\label{def:dei}
The \textit{distributed execution index} (DEI) for an RPC invocation is a sequence $[ r_1|_{c_1} :: r_2|_{c_2} :: \dots :: r_n|_{c_n} ]$ where:

\begin{itemize}[]
    \item $r_n$ is the invocation signature of the invocation; and,
    \item the current RPC invocation is the $c_n$-th invocation of $r_n$ with the path having DEI $[ r_1|_{c_1} :: r_2|_{c_2} :: \dots :: r_{n-1}|_{c_{n-1}} ]$.
\end{itemize}

The definition of a DEI is thus recursive, with the base case being the top-level entry point to the application, whose path is the empty sequence $[~]$.
\end{definition}

\subsection{Special Cases of \ContributionAbbr{}}
\label{sec:dei:specialcases}
 We now look at both \ThreeMB{}~\cite{10.1145/3472883.3486986} and \Filibuster~\cite{10.1145/3472883.3487005} and show how the RPC identification used by both of these systems are special cases of of \ContributionAbbr{}.

\ThreeMB{}~\cite{10.1145/3472883.3486986} uses a combination of signature and invocation count to identify each RPC.  
Rather than track the path of enclosing RPCs to the currently invoking RPC, it accumulates the entire history of RPCs that have been issued up to the current RPC.  
This type of identification is path sensitive, where each RPC is identified using the history of all of the RPCs previously issued before it, and in the same execution order.  
As demonstrated, this is problematic for branching control flow, during exhaustive search, and scheduling nondeterminism, when concurrency is present.

\Filibuster~\cite{10.1145/3472883.3487005} only supports HTTP RPC's and uses a combination of signature, invocation count, and call stack to identify each RPC uniquely.
Without consideration of the payload, this type of identification is problematic for
scheduling nondeterminism, when concurrency is present.
This makes sense: \Filibuster{} only considers single-threaded Python code.


%% file: 04-implementation.tex
\section{Implementation}
\label{sec:implementation}
In order to evaluate \ContributionAbbr{} and demonstrate its applicability to multiple languages and RPC frameworks, we implemented an extension of \Filibuster{} in Java that supports both (A) HTTP-based RPC using the popular Armeria microservice programming framework and (B) Google's popular gRPC framework.  
Both framework choices were influenced by choices made by our industry partner that we plan to use for our evaluation of \ContributionAbbr{} at scale.

Java was chosen as the implementation language for two reasons.
First, Java is a widely used language with both concurrency primitives and true parallelism.
Second, by implementing our algorithm and extensions in Java, they would be able to be used by any language that uses the JVM for its runtime; for example: Scala, Kotlin, and Clojure.
Kotlin specifically is used by the industry partner that we plan to use for our evaluation of \ContributionAbbr{} at scale.

Our work involved the following:

\begin{enumerate}

\item an implementation of the \ContributionAbbr{} algorithm;
\item an implementation of a \Filibuster{} instrumentation library in Java that is used to assign execution indexes to each RPC invocation and maintain the required execution index state for each request; and 
\item several modifications to the existing, open-source \Filibuster{} server (in Python) to generalize its handling of RPC invocations.
\end{enumerate}

This work started in May 2021 and was completed by the end of November 2021.  
In total, our implementation was 11.7 KLOC: 3.4 KLOC of implementation code with 8.3 KLOC of test code.

\subsection{Instrumentation Challenges}

Modifications to the existing, open-source \Filibuster{} prototype to generalize its tracking of RPC invocations for exhaustive search were straightforward: generalizing HTTP as a RPC invocation with a target, method, and parameters.  
Similarly, implementation of the \ContributionAbbr{} algorithm and associated data structures also straightforward.
However, propagation of \ContributionAbbr{} state -- required to construct the concatenative execution indexes that identify RPCs deep in the graph (\textit{i.e.,} identifying the B to E RPC, when A calls B which causes B to call E) -- proved quite challenging in practice. 
Most of these challenge arose from internal thread management, typically done to improve performance, in the underlying web or RPC framework library code used to issue or handle RPCs.

Consider \textsc{DeadlockFuzzer}, a Java tool for the identification of deadlocks using execution indexing.
\textsc{DeadlockFuzzer} instruments the Java byte code at compile time to control scheduling and give it visibility into thread creation and synchronization.  
Then, thread-local state is used to track the internal state required for each execution index.
In contrast, \Filibuster{} is trying to avoid scheduler control (or  specialized compilation for testing) by crafting execution indexes that distinguish concurrent RPCs.
This is required at the scale of microservice applications and their potential language diversity.

The implications of this are that thread-local state \textit{used naively} is not sufficient, unless you have visibility into Java thread operations (\textit{e.g.,} creation, context switching, etc.)
This is essential given the common performance optimizations seen in web and RPC frameworks today. 

We provide a few demonstrative examples of concrete challenges we encountered when implementing various prototype solutions:

\begin{itemize}
	
	\item 
	\textit{Event Loops} are found in the implementations of many microservice frameworks.  
	Used for single-threaded server implementations, they are never descheduled to avoid the penalty of context switches and handle incoming RPCs both serially and synchronously.  Therefore, developers are encouraged (and, in many cases forced through runtime exceptions) to perform any asynchronous operations using futures on a specific thread pool.  
	When the user has to perform blocking operations, they are encouraged to use a specific thread pool for blocking operations, in order to avoid scheduling issues with other nonblocking operations, or deadlocks with other blocking operations that might compete for a shared resource using locks, mutexes, semaphores, or introduce cycles into the microservice graph (\textit{i.e.,} service reentrancy.)  
	Armeria is one example of a microservice framework that uses an event loop.    

	\item 
	\textit{Thread Pools} pose problems when using thread local variables due to thread reuse: thread local state will either be uninitialized or contain values from work performed by the last function executing on that thread.  
	For example, in Java, developers are able to create \texttt{Runnable} objects containing code that should be executed on a different thread.  
	When executing these objects, developers are able to specify a thread pool to use for execution; if not specified, a shared ``common pool'' of threads supplied by the JVM are used.
	
	\item 
	\textit{Asynchronous IO} is a feature provided by most RPC frameworks.  These frameworks are typically implemented with and share the same problems as thread pools: a single RPC may execute across different threads.  
	For example, an RPC between two services may start on one thread when issuing an RPC, be suspended while waiting for IO, and then resume execution on a different thread.  
		
	\item 
	\textit{Futures}, common in asynchronous code implemented in Java, can contain arbitrary user code and execute on either a developer specified thread pool or the common thread pool. (\textit{e.g.,} \mintinline{java}{CompletableFuture})  
	
	\item 
	\textit{Coroutines} and other types of suspendable functions, such as those provided by languages that compile to Java (\textit{e.g.,} Kotlin, Scala), may execute across several different threads during their execution and without allowing the user to specify a custom thread pool for their execution.  	
\end{itemize}

To address this problem, we researched open-source distributed tracing frameworks to understand how they addressed this problem.
This research led us to the OpenTelemetry project, used to automatically instrument many libraries used when building microservice applications to enable distributed tracing.
OpenTelemetry achieves using a combination of three different techniques, of which we both leveraged for our \Filibuster{} integration.

\begin{enumerate}

\item 
Applications are not directly modified for instrumentation.  
Instead, a runtime parameter \mintinline{java}{-javaagent} is supplied that points to a JAR file containing code that is allowed to instrument libraries at runtime with arbitrary code using an API provided by Java.
\item 
Using this API, the standard Java concurrency libraries (and standard libraries for other languages that run on the JVM) are installed that allows OpenTelemetry to automatically migrate context information between threads when context switches occur (or, other concurrency mechanisms are present, such as coroutines where values must be moved between coroutine scopes.)
\item
Finally, this same API is used to install code, specific to each web framework, RPC framework, and database client supported by OpenTelemetry, to perform the distributed tracing using the available context information from (2).
This is the API that we leveraged to automatically install the required \Filibuster{} instrumentation and integrate our client library for Java with the proper context information needed for tracking and propagating our execution indexes.

\end{enumerate}

\subsection{Streaming Challenges}
Including the payload in in the execution index was also not a straightforward engineering task either.
This is a result of many RPC frameworks supporting streaming and being structured, in their implementation, streaming-\textit{first}: where, even when a single message is sent, it is structured as a stream containing a single message.

This proved challenging for \Filibuster{}.  
For example, \Filibuster{} propagates execution indexes between different services using headers or protocol-specific metadata. 
This is in direct contrast to \ThreeMB{}, which modifies the protocol definitions between services and assigns metadata inside of the (de-)serializer to new fields that are used only for fault injection and tracing.
With streams, regardless of the number of messages that are transmitted, headers are transmitted \textit{prior to} the payload, and therefore the payload is not known at the time the headers are transmitted.

For unary messages -- streams containing a single message only -- we were able to buffer the header transmission inside of the \Filibuster{} instrumentation code until the payload was known.  
Once known, we are able to replay the header messages containing the payload encoding as part of the execution index.
However, this does not address the general problem of stream usage.

\subsubsection{Execution Index Rewriting}
To address this problem more generally, we leveraged a property of our execution indexes where they are always unique \textit{for a given execution} without payload inclusion, however, in the presence of scheduling nondeterminism as a result of concurrency, \textit{are only deterministic across executions} when the payload is included.
Therefore, even if we assign incorrect -- but unique -- execution indexes during a single test execution, they can be rewritten to be deterministic as long as it is completed \textit{prior to} both (A) completion of the current test execution and (B) when subsequent test executions are scheduled based on newly discovered program paths.
This remapping from \textit{preliminary} execution index to the actual execution index is performed by the \Filibuster{} server when the RPC is completed once the actual execution index is known.

To demonstrate, consider Figure~\ref{fig:streaming}.  
In this example, scheduling nondeterminism can result in either the \texttt{Hello} RPC executing first and the \texttt{World} RPC executing second, or the reverse where the \texttt{World} RPC executes first and the \texttt{Hello} RPC executes second. 
Regardless of the execution order, the RPC signature and call stack will be the same for both RPCs; the only difference in these RPCs is the payload.  

\begin{figure}[t]
\begin{minted}[obeytabs=true,tabsize=4,linenos,numbersep=-10pt,fontsize=\footnotesize]{python}
    @service_a.route("/")
    def service_a_index():
        b_stream = create_stream(service_b)
        
        def call_b(string):
            return rpc(b_stream, "/", string)
        words = ["Hello", "World"]
        futures = []
        for i in words:
            futures.append(async call_b(i))
        return await_all(futures).join(" ")
    
    @service_b.route("/")
    def service_b_hello():
        return payload.get_string()
\end{minted}
\caption{Use of RPC streaming API that exhibits scheduling nondeterminism.}
\label{fig:streaming}
\end{figure}

When the stream is opened (Figure~\ref{fig:streaming}, line 3), a preliminary execution index, which uses an empty payload and the location of the stream creation as the call stack when generating the invocation signature, is generated and transmitted to the \Filibuster{} server.
That is represented as $([A(\mintinline{java}{null})^3|_{x}])$  (where $x = 1$ for Figure~\ref{fig:streaming}, specifically.)
We also include a flag in the headers or metadata to indicate that this execution index is preliminary, to distinguish from non-streaming RPCs with empty (or \mintinline{java}{null}) payloads.

When the RPCs are actually invoked at the caller, our instrumentation records the final execution index locally along with a mapping from the preliminary execution index and what item it was on the stream.
This includes the correct invocation signature (containing the correct call stack of the actual location where the message was sent; in Figure~\ref{fig:streaming}, line 6) and the RPC payload.
For \mintinline{java}{Hello}, this is $([A(\mintinline{java}{Hello})^6|_1])$; for \mintinline{java}{World}, this is $([A(\mintinline{java}{World})^6|_1])$.

When the invocation is received by the callee, it implicitly increments the invocation count from the starting, preliminary execution index that was received in the header.  
This incremented execution index is the execution index that is propagated when subsequent RPCs are issued from the callee.
For example, the \nth{1} message containing \mintinline{java}{Hello} is implicitly assigned $([A(\mintinline{java}{null})^3|_{x+1}])$; the \nth{2} message \mintinline{java}{World} is implicitly assigned $([A(\mintinline{java}{null})^3|_{x+2}])$.
These identifiers may be permuted across executions but it does not matter: for \textit{this specific test execution}, they are unique.

When the invocation is complete at the caller, and a response is received for each of these RPCs, the \Filibuster{} server is notified of this completion by the caller with both the preliminary and actual execution indexes.
From there, it rewrites any execution index matching (or containing, if a nested request) these preliminary execution indexes to contain the finalized execution indexes.
It is at this point, the \Filibuster{} server has execution indexes that are both unique and deterministic.

As stated above, all that is necessary is that these \ContributionAbbr{}'s are corrected before the next execution for fault injection is scheduled to ensure a valid execution is being scheduled and that the execution is not redundant with respect to the exhaustive search and any optimizations that rely on these identifiers.
To account for this, we adapted the \Filibuster{} server to delay scheduling additional test executions until all preliminary \ContributionAbbr{} were finalized.
As a note, while we have implemented this solution and tested it for single element streams, we have not evaluated it for larger streams yet.

%% file: 05-evaluation.tex
\section{Preliminary Evaluation}
\label{sec:evaluation}
Using \Filibuster{} and its corpus, we demonstrate the need for the inclusion of invocation count, call stacks, and RPC path into the invocation's identifiers.  
Using our extension of \Filibuster{}, we demonstrate the problem of scheduling nondeterminism and show that the inclusion of the payload avoids these issues.

\begin{figure}
  \includegraphics[width=\linewidth]{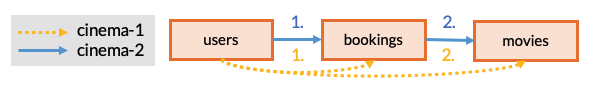}
  \caption{Structure of \textit{cinema-1} and \textit{cinema-2}.}
  \label{fig:evaluation:structure}
\end{figure}

\subsection{Required for Correctness: \\
            Invocation Count, Stack, and Path}
\label{sec:evaluation:corpus}
To demonstrate the need for invocation count, call stacks, and RPC path, we use the \Filibuster{} corpus.
This corpus is composed of 4 industrial examples, re-created from the descriptions of actual microservice applications taken from presentations at industrial conferences on resilience engineering, and 8 ``cinema'' examples, that are used to demonstrate a particular microservice RPC pattern with the help of a microservice application that tracks user's cinema reservations, as briefly described in the introduction.
As none of the industrial examples contained the coding patterns discussed in Section~\ref{sec:dei}, we used the cinema examples in our evaluation. 

With regard to the cinema examples, we were able to identify one example, cinema-3, that demonstrated the need for inclusion of the call stack or invocation count. 
To demonstrate the need for both, we needed to combine the structure of cinema-6 with the use of retries on failure from cinema-1.
We also had to extract the RPC invocation into a helper function.
To demonstrate the need for inclusion of the execution path, we needed to combine the structure of cinema-2 with the use of default responses on failure from cinema-5.
We refer to these \textit{new} examples as cinema-9 and 10, respectively.

Cinema is composed of 4 services, depicted in Figure~\ref{fig:evaluation:structure}.
Users retrieves the bookings for a user: this involves an RPC to bookings and then to movies for each booking.
In the two variations we use, either: a.) users RPCs to movies after the response from bookings; or b.) bookings RPCs to movies directly.
We use a single functional test that get the bookings for a user.  
For fault injection, we consider a single connection error exception per RPC.

Table~\ref{tab:evaluation:table} summarizes these results.

\begin{table}
\begin{center}
\begin{tabular}{|| c || c | c | c | c | c ||} 
 \hline
 Cinema  & All  & No & No    & No IC, & No Path, IC \\
 Ex.     &      & IC & Stk   & Stk    & \& Stk \\
 \hline\hline
 3 & 7 & \textbf{4} & 7 & - & - \\
 \hline
 9 & 5 & 5 & 5 & \textbf{3} & - \\
 \hline
 10 & 6 & 6 & 6 & 6 & \textbf{5}\\
 \hline
\end{tabular}
\end{center}
\caption{Results that demonstrate all techniques must be combined for correct RPC identification.}
\label{tab:evaluation:table}
\end{table}

\paragraph{Invocation count.}
For this experiment, we use cinema-3, where the RPC from users to bookings is done with a loop and re-executed once on failure.
Exhaustive search requires 7 executions; without invocation count, we run 4 executions.

\begin{mdframed}[skipabove=0.1cm, skipbelow=0.1cm, nobreak]
\SFITAbbr{} is incorrect without \textit{invocation counting:}
\begin{itemize}[leftmargin=*,noitemsep,topsep=0pt]
    \item \textit{Unsound.} 
    As each RPC in the loop will be assigned the same identifier, RLFI will either inject a fault on zero or all iterations.
    \item \textit{Incomplete.} 
    Requests that occur as a result of any iteration, not the \nth{1}, will not have faults injected.
\end{itemize}
\end{mdframed}

\paragraph{Call stack.}
For this experiment, we used cinema-9.
In the event of failure of the \nth{1} RPC from users to bookings, it will mark the request as failed and try that request later from a different call site.
This differs from the loop where the same call site is used.  
Each call site uses a helper function to issue the RPC to ensure the stacks are different.
Exhaustive search requires 5 executions; without call stack, we run 3 executions.

\begin{mdframed}[skipabove=0.1cm, skipbelow=0.1cm, nobreak]
\SFITAbbr{} is incorrect without \textit{call stack inclusion:}
\begin{itemize}[leftmargin=*,noitemsep,topsep=0pt]
    \item \textit{Unsound.} 
    As each call invocation of the helper's RPC will be assigned the same identifier, RLFI will either inject a fault on both or neither.
    \item \textit{Incomplete.} 
    Requests that occur as a result of the \nth{1} invocation will not have faults injected.
\end{itemize}
\end{mdframed}

\paragraph{RPC Path.} 
For this experiment, we used cinema-10.
In the event of a failure of bookings, a default response is used in place of the failure and the movies service contacted by the users service directly.
Exhaustive search requires 6 test executions; without RPC path, we run 5 executions.

\begin{mdframed}[skipabove=0.1cm, skipbelow=0.1cm, nobreak]
\SFITAbbr{} is incorrect without \textit{RPC path inclusion:}
\begin{itemize}[leftmargin=*,noitemsep,topsep=0pt]
    \item \textit{Unsound.} 
    Since the bookings RPC to movies and the users RPC to movies share the same identifier, RLFI will either inject a fault on both or neither.  
    \item \textit{Incomplete.} 
    As RLFI will always fail the \nth{2} RPC to movies, we do not explore the successful case.
\end{itemize}
\end{mdframed}

\subsection{Nondeterminism is a Problem}
\label{sec:evaluation:nondeterminism}
In order to understand the impact of scheduling nondeterminism within the JVM on correct identifier assignment, we constructed a small example with the Armeria that contained two services: \textit{Hello} and \textit{World}. 

In this example, the \textit{World} service exposed a single endpoint that returned a \texttt{String} constant when it received an RPC.  
The \textit{Hello} service exposed an endpoint that, when it received an RPC from our test harness, would launch a configurable number of threads, each that issued an RPC to the \textit{World} service, and then wait for them to complete.
Each thread was defined as a class in Java, where the \textit{Hello} service would create instances of this class of in a loop: this ensured that the call site of the RPC was the same and the stack trace of the call site were identical.  
All RPCs were made the same service and differed only in the payload, which contained the identifier of the thread determined by thread creation order.

\begin{figure}
  \includegraphics[width=\linewidth]{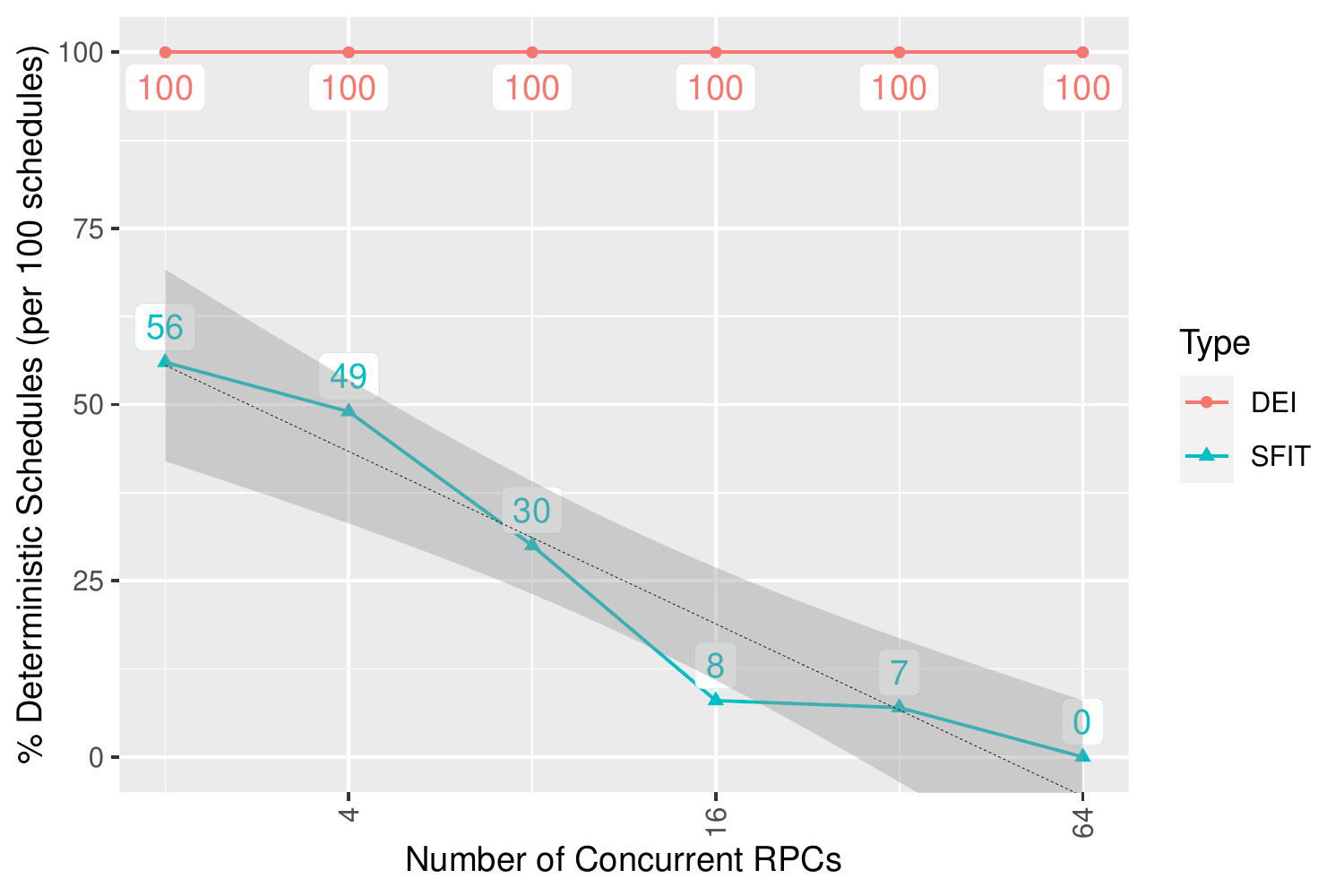}
  \caption{Percentage of executions with deterministic assignment for configuration of 2 threads.}
  \label{fig:evaluation:scheduling-nondeterminism}
\end{figure}

For this experiment, we used our \Filibuster{} implementation extended with support for Java, gRPC, and \ContributionAbbr{}.  
We reconfigured the \ContributionAbbr{} algorithm to include the thread creation order.
Therefore, payload differed only by this identifier.
We ran this test application for varying numbers of RPCs (2, 4, 8, 16, 32, 64) for 100 iterations each.  
We fixed the thread pool size at the \textit{Hello} service at size 2.
For each iteration, we recorded whether or not the execution index assignment matched the thread creation order by examining the execution indexes payload values.
The results are presented in Figure~\ref{fig:evaluation:scheduling-nondeterminism}. 
With only 2 RPCs, 44\% of the tests exhibited an RPC execution order that did not match the creation order; by 64 , 100\% of the RPCs did not match.

\begin{mdframed}[skipabove=0.1cm, skipbelow=0.1cm, nobreak]
\cmark~Even in the presence of relatively low amounts of concurrency, scheduling nondeterminism is a problem for existing RLFI techniques.
\end{mdframed}

\subsection{Payload Inclusion Distinguishes}
\label{sec:evaluation:payload}
Using the same example from Section~\ref{sec:evaluation:nondeterminism}, we were able to verify our \textit{key insight}: inclusion of the payload into the identifier for each RPC invocation was sufficient for distinguishing these concurrent, inter-service RPCs.  
With \ContributionAbbr{}, all RPC identifiers were unique and deterministic, as shown in Figure~\ref{fig:evaluation:scheduling-nondeterminism}.
Therefore, scheduling nondeterminism was not an issue.

\begin{mdframed}[skipabove=0.1cm, skipbelow=0.1cm, nobreak]
\cmark~Payload is sufficient for distinguishing concurrent, inter-service RPCs in microservice applications, under the assumption that these RPCs will not share the same payload when issued to the same service and method with the same parameters.
\end{mdframed}

%% file: 06-conclusion.tex
\section{Related Work}
\label{sec:related-work}

Execution indexing~\cite{10.1145/1379022.1375611} was originally devised to identify unique points in an execution, in a manner that established a correspondence across multiple executions.
It has also been used for deadlock identification~\cite{10.1145/1543135.1542489} and traversal analysis~\cite{10.1109/ICSE.2017.50}.

In Section~\ref{sec:dei:payload}, we described how model checkers for distributed systems~\cite{267763, 10.5555/2685048.2685080, 10.1145/3302424.3303986} were originally designed for identifying concurrency bugs and later extended to test failures.
Therefore, they rely on control of the thread scheduler, which is unrealistic for large, microservice applications that run on hundreds of different machines, implemented in different languages.

In Section~\ref{sec:dei:specialcases}, we discussed the differences between \Filibuster{}~\cite{10.1145/3472883.3487005, filibuster} and \ThreeMB{}~\cite{10.1145/3472883.3486986}, two modern approaches to RLFI, and demonstrated that the techniques used by both are special cases of \ContributionAbbr{}.  

Many practitioners still rely on stochastic techniques~\cite{8383672, chaos-monkey, gremlin, alfi, 10.1109/ICSE-SEIP.2019.00012} that attempt to minimize the blast radius of random experiments in production: the \Filibuster{} paper identifies 32 different companies all using this style of experimentation on microservice applications.
Recently, there has been interest in using these techniques into the local development environment~\cite{chaos-toolkit, linkedout, 7536505} to minimize the blast radius even further.
However, these techniques each require that developers manually specify the fault configurations that are tested, necessitating the need for a mechanism such as \ContributionAbbr{}, that supports a sound and complete systematic search.

\section{Future Work}
In this section, we present several ideas for future work.

\subsection{Alternative Instantiations}
We believe that a promising area of research is in exploring different instantiations of the \contribution{} algorithm.  

As an example, we have already shown that two different instantiations have been of value.
For example, \Filibuster{}'s instantiation, which supports fault injection and exhaustive search in applications that contain no concurrency and only use synchronous RPC.
Similarly, \ThreeMB{}'s instantiation, which provides temporal, nondeterministic concurrent fault injection in a live system.
We envision several different instantiations that may be of use for microservice application developers.

First, we believe that an instantiation that does not include (or more precisely sets to \mintinline{java}{null}) the invocation count or stack trace may be useful in identifying a particular microservice anti-pattern where the same RPC endpoint is accessed by the same service multiple times as part of a single request.
We have empirically observed this pattern several times.
For example, a service contains a helper function called \mintinline{java}{getCurrentUser} that is repeatedly called at different locations by a service as part of processing a single request instead of retrieving the user a single time and passing those values around through function calls.  
We hypothesize that this pattern emerges from the combination of abstraction and the lack of visibility where RPC's actually occur within an application and a result of this, a perceived lack of cost.

We also believe that same instantiation, augmented by an analysis, can be used to detect a second microservice anti-pattern: specifically, where a remote services does not directly expose a required API and therefore the invoker of those services must combine two remote calls to get the information they require.
For example, calling \mintinline{java}{getCurrentUser} and then immediately calling \mintinline{java}{getProfileByUser} and providing the user record that was returned from the first call as an argument to the second.
We hypothesize that these patterns arise from the decentralized nature of microservice development where a single developer of a service must implement their functionality with ``what's available'' as APIs from the services that they take dependencies on.  

Finally, we believe that when testing more advanced resilience techniques such as circuit breakers and load shedding, other instantiations will be necessary.
When it comes to circuit breakers that operate on \textit{per service}, it may be useful to perform fault injection on any execution index that targets a particular service.  
This may require coarser execution indexes that do not discriminate for invocation count, stack trace, or payload if the goal is to inject a fault for each RPC to that service.
Similarly, the same might apply to load shedding as well.
For circuit breakers that operate \textit{per method}, a different instantiation may be required that includes the stack trace, or a minimized stack trace that only considers the final frame.  
For load shedding that operates \textit{per method}, an instantiation that only considers the RPC signature maybe appropriate.

Regardless of the instantiation and the precise use case, we believe that the \contribution{} algorithm provides a framework for supporting all of these and identifying the mapping between instantiations and applications is a promising future research direction.
In fact, we have already begun work on several of these and a mechanism for projecting executions indexes with full instantiation into execution indexes containing these alternative instantiations at runtime for analysis without requiring that the system generate and maintain all possible representations.

\subsection{Graph Analysis}
One interesting property of the execution indexes produced by the \contribution{} algorithm is that they can be used to determine the structure of the application graph.  
For example, if Service A calls Service B and Service B calls Service C, the execution index for the specific call between B and C will contain the execution index of the call from A to B as a prefix.  
Not only can this be used to dynamically reconstruct the application graph from recorded observations of traces --- for example, if storing the augmented OpenTelemetry information in Jaeger --- but we believe that this information could also be used for detection and activation of fault tolerance mechanisms at runtime.

One such application is to detect when circuit breakers have been activated. 
For example, by injecting faults repeatedly on a request with a given execution index and subsequently observing the disappearance of that execution index in a trace, along with all execution indexes that contain that execution index a a prefix, in subsequent traces, one might be able to conclude that a circuit breaker opened and prevented subsequent requests to the request where the fault was injected.  
By continuing to issue requests where no faults are injected and subsequently observing the subgraph restored in following traces, one may be able to conclude that the circuit breaker has re-closed.

The key here is that the execution indexes are stable under control flow changes, so when the circuit breaker opens, any subsequent requests that occur are guaranteed to not appear like or be confused with the requests that occurred when the circuit breaker was closed.
This is one possible application of \contribution{} for identifying fault tolerance mechanisms through observation of traces containing execution indexes.

\subsection{Implementation}
When it comes to implementation, our \Filibuster{} extension still needs to be evaluated at scale, with different concurrency primitives, to determine it's viability.

For example, when it comes to Kotlin specifically, we identified a several situtation where, when using suspendable functions combined with Armeria's futures -- used when issuing both HTTP and GRPC calls -- the call stack did not contain any frames originating in the application code.   
This makes identification of the actual RPC location in application code not possible.
We believe this to be an artifact of Kotlin's coroutine handling and suspect that deeper integration with Kotlin's stacktrace reassembly mechanism may be necessary to properly identify the location where the RPCs are actually invoked in the application code.

Somewhat similarly, Kotlin coroutines may be subject to multiple context switches and rescheduling across different threads.  
The result of this is that, depending on scheduling nondeterminism, call stacks may differ across executions by one or more frames originating in internal language libraries (\textit{e.g.,} \mintinline{java}{java.lang.Thread}) thereby explicitly encoding the  context switches or other scheduling decisions into the call stack itself.
To address this, we implemented a mechanism for white/blacklisting frames in the call stack related to standard libraries.
This has proven useful in preliminary tests, but has not been evaluated at scale yet.

\subsection{Algorithm}
Inclusion of the RPC signature, call stack, and invocation path -- what can be thought of as \textit{synchronous} \contribution{} -- has already been evaluated as part of our work on \SFITAbbr{} and was published at ACM's Symposium on Cloud Computing in 2021~\cite{10.1145/3472883.3487005}.
However, while evaluated synthetically, \textit{asynchronous} \contribution{} -- where the payload is included in the execution index -- still needs further evaluation in an industrial microservice application in order to demonstrate its viability.

The design decisions that underly the inclusion of the payload in the execution index are rooted in observations we made when examining a large-scale, industrial microservice application, composed of over 500 different services.

Examination of this application resulted in two \textit{key observations:}

\begin{enumerate}
	\item 
	In the majority of cases where concurrent RPCs were executed, asynchronously, to the same service and method, from the same call site, they were typically done to retrieve \textit{different} information in parallel, as fast as possible, as indicated by the arguments to the RPC. 
	For example, retrieving records from a database and then issuing RPCs inside of a loop to aggregate those records.
	There is some risk here: databases without proper constraints may return a list of records containing duplicate entries resulting in concurrent retrieval of the same record.  
	\item
	In a few of the cases we identified, parallel execution of the same workflow, parameterized by the same arguments, resulted in concurrent execution of the same \textit{single} RPC invocation.
\end{enumerate}

These key observations resulted in the following \textit{key insight.}
When concurrent RPCs are executed, asynchronously, to the same service and method, from the same call site for the same payload, for the purposes of aggregation, permutation of the identifiers for these RPCs is observationally equivalent.  
In that, whether we inject a fault on the first or second has no outcome on the program, as both responses from each RPC will be the same. 
This is true for any API that is true for retrieving records, but may not hold true for APIs that perform mutations.  
When it comes to mutations, this holds true for any mutation that is deterministic, idempotent, and commutative: and therefore may apply to some, but not all.
	
Further evaluation is still required in order to understand if these observations hold true across all industrial microservice applications.

\section{Conclusion and Future Work}
\label{sec:conclusion}
In this paper, we presented \textit{\contribution{}} (\ContributionAbbr{}), a technique for microservice applications that precisely identifies dynamic instances of inter-service RPCs.
\ContributionAbbr{} addresses a real need in modern, microservice resilience testing, as existing RLFI techniques all fail to handle common RPC patterns that exist in industrial microservice applications.
We formally defined the general concept of \ContributionAbbr{} and demonstrated that two of the most recent RLFI systems use special cases of \ContributionAbbr{}.

%% file: 00-main.bbl

\begin{thebibliography}{19}


\ifx \showCODEN    \undefined \def \showCODEN     #1{\unskip}     \fi
\ifx \showDOI      \undefined \def \showDOI       #1{#1}\fi
\ifx \showISBNx    \undefined \def \showISBNx     #1{\unskip}     \fi
\ifx \showISBNxiii \undefined \def \showISBNxiii  #1{\unskip}     \fi
\ifx \showISSN     \undefined \def \showISSN      #1{\unskip}     \fi
\ifx \showLCCN     \undefined \def \showLCCN      #1{\unskip}     \fi
\ifx \shownote     \undefined \def \shownote      #1{#1}          \fi
\ifx \showarticletitle \undefined \def \showarticletitle #1{#1}   \fi
\ifx \showURL      \undefined \def \showURL       {\relax}        \fi
\providecommand\bibfield[2]{#2}
\providecommand\bibinfo[2]{#2}
\providecommand\natexlab[1]{#1}
\providecommand\showeprint[2][]{arXiv:#2}

\bibitem[\protect\citeauthoryear{??}{lin}{2018}]%
        {linkedout}
 \bibinfo{year}{2018}\natexlab{}.
\newblock \bibinfo{title}{{LinkedOut: A Request-Level Failure Injection
  Framework}}.
\newblock
  \bibinfo{howpublished}{\url{https://engineering.linkedin.com/blog/2018/05/linkedout--a-request-level-failure-injection-framework}}.
\newblock
\newblock
\shownote{Accessed: 2021-05-21.}


\bibitem[\protect\citeauthoryear{??}{cha}{2021a}]%
        {chaos-monkey}
 \bibinfo{year}{2021}\natexlab{a}.
\newblock \bibinfo{title}{{Chaos Monkey}}.
\newblock \bibinfo{howpublished}{\url{https://netflix.github.io/chaosmonkey/}}.
\newblock
\newblock
\shownote{Accessed: 2021-11-30.}


\bibitem[\protect\citeauthoryear{??}{cha}{2021b}]%
        {chaos-toolkit}
 \bibinfo{year}{2021}\natexlab{b}.
\newblock \bibinfo{title}{{Chaos Toolkit}}.
\newblock \bibinfo{howpublished}{\url{https://chaostoolkit.org/}}.
\newblock
\newblock
\shownote{Accessed: 2021-11-30.}


\bibitem[\protect\citeauthoryear{??}{fil}{2021}]%
        {filibuster}
 \bibinfo{year}{2021}\natexlab{}.
\newblock \bibinfo{title}{{Filibuster}}.
\newblock \bibinfo{howpublished}{\url{http://filibuster.cloud}}.
\newblock
\newblock
\shownote{Accessed: 2021-09-07.}


\bibitem[\protect\citeauthoryear{??}{alf}{2021}]%
        {alfi}
 \bibinfo{year}{2021}\natexlab{}.
\newblock \bibinfo{title}{{Getting Started with Application Level Failure
  Injection}}.
\newblock \bibinfo{howpublished}{\url{https://bit.ly/3pg8m8k}}.
\newblock
\newblock
\shownote{Accessed: 2021-11-30.}


\bibitem[\protect\citeauthoryear{??}{gre}{2021}]%
        {gremlin}
 \bibinfo{year}{2021}\natexlab{}.
\newblock \bibinfo{title}{{Gremlin}}.
\newblock \bibinfo{howpublished}{\url{http://www.gremlin.com}}.
\newblock
\newblock
\shownote{Accessed: 2021-05-21.}


\bibitem[\protect\citeauthoryear{Alvaro, Andrus, Sanden, Rosenthal, Basiri, and
  Hochstein}{Alvaro et~al\mbox{.}}{2016}]%
        {10.1145/2987550.2987555}
\bibfield{author}{\bibinfo{person}{Peter Alvaro}, \bibinfo{person}{Kolton
  Andrus}, \bibinfo{person}{Chris Sanden}, \bibinfo{person}{Casey Rosenthal},
  \bibinfo{person}{Ali Basiri}, {and} \bibinfo{person}{Lorin Hochstein}.}
  \bibinfo{year}{2016}\natexlab{}.
\newblock \showarticletitle{Automating Failure Testing Research at Internet
  Scale}. In \bibinfo{booktitle}{\emph{Proceedings of the Seventh ACM Symposium
  on Cloud Computing}} (Santa Clara, CA, USA) \emph{(\bibinfo{series}{SoCC
  '16})}. \bibinfo{publisher}{Association for Computing Machinery},
  \bibinfo{address}{New York, NY, USA}, \bibinfo{pages}{17–28}.
\newblock
\showISBNx{9781450345255}
\urldef\tempurl%
\url{https://doi.org/10.1145/2987550.2987555}
\showDOI{\tempurl}


\bibitem[\protect\citeauthoryear{Basiri, Hochstein, Jones, and Tucker}{Basiri
  et~al\mbox{.}}{2019}]%
        {10.1109/ICSE-SEIP.2019.00012}
\bibfield{author}{\bibinfo{person}{Ali Basiri}, \bibinfo{person}{Lorin
  Hochstein}, \bibinfo{person}{Nora Jones}, {and} \bibinfo{person}{Haley
  Tucker}.} \bibinfo{year}{2019}\natexlab{}.
\newblock \showarticletitle{Automating Chaos Experiments in Production}. In
  \bibinfo{booktitle}{\emph{Proceedings of the 41st International Conference on
  Software Engineering: Software Engineering in Practice}} (Montreal, Quebec,
  Canada) \emph{(\bibinfo{series}{ICSE-SEIP '19})}. \bibinfo{publisher}{IEEE
  Press}, \bibinfo{pages}{31–40}.
\newblock
\urldef\tempurl%
\url{https://doi.org/10.1109/ICSE-SEIP.2019.00012}
\showDOI{\tempurl}


\bibitem[\protect\citeauthoryear{Heorhiadi, Rajagopalan, Jamjoom, Reiter, and
  Sekar}{Heorhiadi et~al\mbox{.}}{2016}]%
        {7536505}
\bibfield{author}{\bibinfo{person}{Victor Heorhiadi}, \bibinfo{person}{Shriram
  Rajagopalan}, \bibinfo{person}{Hani Jamjoom}, \bibinfo{person}{Michael~K.
  Reiter}, {and} \bibinfo{person}{Vyas Sekar}.}
  \bibinfo{year}{2016}\natexlab{}.
\newblock \showarticletitle{Gremlin: Systematic Resilience Testing of
  Microservices}. In \bibinfo{booktitle}{\emph{2016 IEEE 36th International
  Conference on Distributed Computing Systems (ICDCS)}}.
  \bibinfo{pages}{57--66}.
\newblock
\urldef\tempurl%
\url{https://doi.org/10.1109/ICDCS.2016.11}
\showDOI{\tempurl}


\bibitem[\protect\citeauthoryear{Joshi, Park, Sen, and Naik}{Joshi
  et~al\mbox{.}}{2009}]%
        {10.1145/1543135.1542489}
\bibfield{author}{\bibinfo{person}{Pallavi Joshi}, \bibinfo{person}{Chang-Seo
  Park}, \bibinfo{person}{Koushik Sen}, {and} \bibinfo{person}{Mayur Naik}.}
  \bibinfo{year}{2009}\natexlab{}.
\newblock \showarticletitle{A Randomized Dynamic Program Analysis Technique for
  Detecting Real Deadlocks}.
\newblock \bibinfo{journal}{\emph{SIGPLAN Not.}} \bibinfo{volume}{44},
  \bibinfo{number}{6} (\bibinfo{date}{June} \bibinfo{year}{2009}),
  \bibinfo{pages}{110–120}.
\newblock
\showISSN{0362-1340}
\urldef\tempurl%
\url{https://doi.org/10.1145/1543135.1542489}
\showDOI{\tempurl}


\bibitem[\protect\citeauthoryear{Leesatapornwongsa, Hao, Joshi, Lukman, and
  Gunawi}{Leesatapornwongsa et~al\mbox{.}}{2014}]%
        {10.5555/2685048.2685080}
\bibfield{author}{\bibinfo{person}{Tanakorn Leesatapornwongsa},
  \bibinfo{person}{Mingzhe Hao}, \bibinfo{person}{Pallavi Joshi},
  \bibinfo{person}{Jeffrey~F. Lukman}, {and} \bibinfo{person}{Haryadi~S.
  Gunawi}.} \bibinfo{year}{2014}\natexlab{}.
\newblock \showarticletitle{SAMC: Semantic-Aware Model Checking for Fast
  Discovery of Deep Bugs in Cloud Systems}. In
  \bibinfo{booktitle}{\emph{Proceedings of the 11th USENIX Conference on
  Operating Systems Design and Implementation}} (Broomfield, CO)
  \emph{(\bibinfo{series}{OSDI'14})}. \bibinfo{publisher}{USENIX Association},
  \bibinfo{address}{USA}, \bibinfo{pages}{399–414}.
\newblock
\showISBNx{9781931971164}


\bibitem[\protect\citeauthoryear{Lukman, Ke, Stuardo, Suminto, Kurniawan,
  Simon, Priambada, Tian, Ye, Leesatapornwongsa, Gupta, Lu, and Gunawi}{Lukman
  et~al\mbox{.}}{2019}]%
        {10.1145/3302424.3303986}
\bibfield{author}{\bibinfo{person}{Jeffrey~F. Lukman}, \bibinfo{person}{Huan
  Ke}, \bibinfo{person}{Cesar~A. Stuardo}, \bibinfo{person}{Riza~O. Suminto},
  \bibinfo{person}{Daniar~H. Kurniawan}, \bibinfo{person}{Dikaimin Simon},
  \bibinfo{person}{Satria Priambada}, \bibinfo{person}{Chen Tian},
  \bibinfo{person}{Feng Ye}, \bibinfo{person}{Tanakorn Leesatapornwongsa},
  \bibinfo{person}{Aarti Gupta}, \bibinfo{person}{Shan Lu}, {and}
  \bibinfo{person}{Haryadi~S. Gunawi}.} \bibinfo{year}{2019}\natexlab{}.
\newblock \showarticletitle{FlyMC: Highly Scalable Testing of Complex
  Interleavings in Distributed Systems}. In
  \bibinfo{booktitle}{\emph{Proceedings of the Fourteenth EuroSys Conference
  2019}} (Dresden, Germany) \emph{(\bibinfo{series}{EuroSys '19})}.
  \bibinfo{publisher}{Association for Computing Machinery},
  \bibinfo{address}{New York, NY, USA}, Article \bibinfo{articleno}{20},
  \bibinfo{numpages}{16}~pages.
\newblock
\showISBNx{9781450362818}
\urldef\tempurl%
\url{https://doi.org/10.1145/3302424.3303986}
\showDOI{\tempurl}


\bibitem[\protect\citeauthoryear{Meiklejohn, Estrada, Song, Miller, and
  Padhye}{Meiklejohn et~al\mbox{.}}{2021}]%
        {10.1145/3472883.3487005}
\bibfield{author}{\bibinfo{person}{Christopher~S. Meiklejohn},
  \bibinfo{person}{Andrea Estrada}, \bibinfo{person}{Yiwen Song},
  \bibinfo{person}{Heather Miller}, {and} \bibinfo{person}{Rohan Padhye}.}
  \bibinfo{year}{2021}\natexlab{}.
\newblock \showarticletitle{Service-Level Fault Injection Testing}. In
  \bibinfo{booktitle}{\emph{Proceedings of the ACM Symposium on Cloud
  Computing}} (Seattle, WA, USA) \emph{(\bibinfo{series}{SoCC '21})}.
  \bibinfo{publisher}{Association for Computing Machinery},
  \bibinfo{address}{New York, NY, USA}, \bibinfo{pages}{388–402}.
\newblock
\showISBNx{9781450386388}
\urldef\tempurl%
\url{https://doi.org/10.1145/3472883.3487005}
\showDOI{\tempurl}


\bibitem[\protect\citeauthoryear{Padhye and Sen}{Padhye and Sen}{2017}]%
        {10.1109/ICSE.2017.50}
\bibfield{author}{\bibinfo{person}{Rohan Padhye} {and} \bibinfo{person}{Koushik
  Sen}.} \bibinfo{year}{2017}\natexlab{}.
\newblock \showarticletitle{Travioli: A Dynamic Analysis for Detecting
  Data-Structure Traversals}. In \bibinfo{booktitle}{\emph{2017 IEEE/ACM 39th
  International Conference on Software Engineering (ICSE)}}.
  \bibinfo{pages}{473--483}.
\newblock
\urldef\tempurl%
\url{https://doi.org/10.1109/ICSE.2017.50}
\showDOI{\tempurl}


\bibitem[\protect\citeauthoryear{Tucker, Hochstein, Jones, Basiri, and
  Rosenthal}{Tucker et~al\mbox{.}}{2018}]%
        {8383672}
\bibfield{author}{\bibinfo{person}{H. Tucker}, \bibinfo{person}{L. Hochstein},
  \bibinfo{person}{N. Jones}, \bibinfo{person}{A. Basiri}, {and}
  \bibinfo{person}{C. Rosenthal}.} \bibinfo{year}{2018}\natexlab{}.
\newblock \showarticletitle{The Business Case for Chaos Engineering}.
\newblock \bibinfo{journal}{\emph{IEEE Cloud Computing}} \bibinfo{volume}{5},
  \bibinfo{number}{03} (\bibinfo{date}{may} \bibinfo{year}{2018}),
  \bibinfo{pages}{45--54}.
\newblock
\showISSN{2372-2568}
\urldef\tempurl%
\url{https://doi.org/10.1109/MCC.2018.032591616}
\showDOI{\tempurl}


\bibitem[\protect\citeauthoryear{Xin, Sumner, and Zhang}{Xin
  et~al\mbox{.}}{2008}]%
        {10.1145/1379022.1375611}
\bibfield{author}{\bibinfo{person}{Bin Xin}, \bibinfo{person}{William~N.
  Sumner}, {and} \bibinfo{person}{Xiangyu Zhang}.}
  \bibinfo{year}{2008}\natexlab{}.
\newblock \showarticletitle{Efficient Program Execution Indexing}.
\newblock \bibinfo{journal}{\emph{SIGPLAN Not.}} \bibinfo{volume}{43},
  \bibinfo{number}{6} (\bibinfo{date}{June} \bibinfo{year}{2008}),
  \bibinfo{pages}{238–248}.
\newblock
\showISSN{0362-1340}
\urldef\tempurl%
\url{https://doi.org/10.1145/1379022.1375611}
\showDOI{\tempurl}


\bibitem[\protect\citeauthoryear{Yang, Chen, Wu, Xu, Liu, Lin, Yang, Long,
  Zhang, and Zhou}{Yang et~al\mbox{.}}{2009}]%
        {267763}
\bibfield{author}{\bibinfo{person}{Junfeng Yang}, \bibinfo{person}{Tisheng
  Chen}, \bibinfo{person}{Ming Wu}, \bibinfo{person}{Zhilei Xu},
  \bibinfo{person}{Xuezheng Liu}, \bibinfo{person}{Haoxiang Lin},
  \bibinfo{person}{Mao Yang}, \bibinfo{person}{Fan Long},
  \bibinfo{person}{Lintao Zhang}, {and} \bibinfo{person}{Lidong Zhou}.}
  \bibinfo{year}{2009}\natexlab{}.
\newblock \showarticletitle{{MODIST}: Transparent Model Checking of Unmodified
  Distributed Systems}. In \bibinfo{booktitle}{\emph{6th {USENIX} Symposium on
  Networked Systems Design and Implementation ({NSDI} 09)}}.
  \bibinfo{publisher}{{USENIX} Association}, \bibinfo{address}{Boston, MA}.
\newblock
\urldef\tempurl%
\url{https://www.usenix.org/conference/nsdi-09/modist-transparent-model-checking-unmodified-distributed-systems}
\showURL{%
\tempurl}


\bibitem[\protect\citeauthoryear{Yuan, Luo, Zhuang, Rodrigues, Zhao, Zhang,
  Jain, and Stumm}{Yuan et~al\mbox{.}}{2014}]%
        {186171}
\bibfield{author}{\bibinfo{person}{Ding Yuan}, \bibinfo{person}{Yu Luo},
  \bibinfo{person}{Xin Zhuang}, \bibinfo{person}{Guilherme~Renna Rodrigues},
  \bibinfo{person}{Xu Zhao}, \bibinfo{person}{Yongle Zhang},
  \bibinfo{person}{Pranay~U. Jain}, {and} \bibinfo{person}{Michael Stumm}.}
  \bibinfo{year}{2014}\natexlab{}.
\newblock \showarticletitle{Simple Testing Can Prevent Most Critical Failures:
  An Analysis of Production Failures in Distributed Data-Intensive Systems}. In
  \bibinfo{booktitle}{\emph{11th {USENIX} Symposium on Operating Systems Design
  and Implementation ({OSDI} 14)}}. \bibinfo{publisher}{{USENIX} Association},
  \bibinfo{address}{Broomfield, CO}, \bibinfo{pages}{249--265}.
\newblock
\showISBNx{978-1-931971-16-4}
\urldef\tempurl%
\url{https://www.usenix.org/conference/osdi14/technical-sessions/presentation/yuan}
\showURL{%
\tempurl}


\bibitem[\protect\citeauthoryear{Zhang, Ferydouni, Montana, Bittman, and
  Alvaro}{Zhang et~al\mbox{.}}{2021}]%
        {10.1145/3472883.3486986}
\bibfield{author}{\bibinfo{person}{Jun Zhang}, \bibinfo{person}{Robert
  Ferydouni}, \bibinfo{person}{Aldrin Montana}, \bibinfo{person}{Daniel
  Bittman}, {and} \bibinfo{person}{Peter Alvaro}.}
  \bibinfo{year}{2021}\natexlab{}.
\newblock \showarticletitle{3MileBeach: A Tracer with Teeth}. In
  \bibinfo{booktitle}{\emph{Proceedings of the ACM Symposium on Cloud
  Computing}} (Seattle, WA, USA) \emph{(\bibinfo{series}{SoCC '21})}.
  \bibinfo{publisher}{Association for Computing Machinery},
  \bibinfo{address}{New York, NY, USA}, \bibinfo{pages}{458–472}.
\newblock
\showISBNx{9781450386388}
\urldef\tempurl%
\url{https://doi.org/10.1145/3472883.3486986}
\showDOI{\tempurl}


\end{thebibliography}
